\documentclass[a4paper,10pt]{article}

\title{BRST and Anti-BRST Symmetries in  Perturbative Quantum Gravity}
\author{Mir Faizal\\ Department of Mathematics,  Durham University,\\ Durham, DH1 3LE,  United Kingdom,\\ faizal.mir@durham.ac.uk}

\begin{document}

\maketitle

\begin{abstract}
In perturbative quantum gravity, the sum of the classical Lagrangian density, a gauge fixing term and a ghost term  is invariant under two  sets of supersymmetric
 transformations called the BRST and the anti-BRST transformations. 
In this paper we will analyse the BRST and the anti-BRST symmetries of perturbative quantum gravity in curved spacetime,  
in  linear as well as non-linear gauges. We will show that even though the sum of ghost term and 
the gauge fixing term  can always  be expressed as a total BRST or a total anti-BRST variation, we  can express it as a combination of both of them only 
in certain special gauges. We will also analyse the violation of nilpotency of the BRST and the anti-BRST transformations by 
 introduction of a bare mass term, in the massive Curci-Ferrari gauge. 
\end{abstract}

Key words:   BRST, Anti-BRST, Perturbative quantum gravity 

PACS number: 04.60.-m

\section{Introduction}
Three out of the four fundamental forces in nature are described by Yang-Mills theories. The fourth, being gravity, 
is described by gauge theory of diffeomorphism [1]. In this sense all the forces of nature can be formulated 
in the language  of  gauge theory. 

However, when analysing  with any gauge theory, we have to deal with the redundant degrees of freedom due
 to gauge invariance of that theory. We have to eliminate  these redundant degrees of freedom before  
trying to quantize that theory. An elegant formalism called the BRST formalism is usually employed for this purpose  
[2]. In this formalism the sum of the classical Lagrangian density, a gauge fixing term and a ghost term  
(collectively called a gauge fixing Lagrangian density in this paper) is invariant under a  set of supersymmetric
 transformations called the BRST transformations. This total Lagrangian density is also invariant under another
 set of supersymmetric transformations called the anti-BRST transformations [3]. 

The BRST and the anti-BRST symmetries for perturbative quantum gravity in four dimensional flat spacetime
 have been studied by a number of  authors [4-6] and their work has been summarized by N. Nakanishi and I. Ojima [7].
 The BRST symmetry in two dimensional curved  spacetime has  been thoroughly studied  [8-10].  The BRST and the anti-BRST symmetries 
for topological quantum gravity in  curved spacetime  have also been studied  [11-12]. All this work has been done in linear gauges. 

However BRST and anti-BRST symmetries are known to have a richer structure in Yang-Mills theories. In case of
 Yang-Mills theories, it is known that in Landau gauge we can express the gauge fixing Lagrangian density as 
a combination of  total BRST and total anti-BRST variations [13].  This is also  achieved by  addition of suitable 
 non-linear terms to   the gauge fixing Lagrangian density [14]. Furthermore, the addition of a bare mass 
term breaks the nilpotency of the BRST and the anti-BRST transformations and this leads to the violation of unitarity of the resultant theory [15].

In this paper we will try to generalize these results that are known in the context of Yang-Mills theories in four dimensional 
flat spacetime to perturbative quantum gravity in curved spacetime, in arbitrary dimensions.
 It may be noted that the violation of unitarity did not have much physical relevance in the context of Yang-Mills theories.  
However it is suspected  that certain quantum gravitational processes might lead to violation of  unitarity [16]. So this loss 
of unitarity, due to the addition of a bare mass term, seems to be physically more relevant to quantum gravity in curved spacetime
 than Yang-Mills theories in flat spacetime. 
  
\section{BRST and Anti-BRST Transformations}

The Lagrangian density for pure Euclidean gravity with  cosmological constant $\lambda$ is given by
\begin{equation}
 \mathcal{L} = \sqrt{g}(R - 2\lambda),
\end{equation}
where we have adopted  units, such that
\begin{equation}
 16 \pi G = 1.
\end{equation}
In perturbative gravity one splits the full metric $g_{ab}^f$ into the metric for the
 background spacetime $g_{ab}$ and a small perturbation around it being $h_{ab}$. 
The covariant derivatives along with the lowering and raising of indices are compatible with the 
metric for the background spacetime. The small perturbation $h_{ab}$ is viewed as the field that is to be quantized.

All the degrees of freedom in $h_{ab}$ are not physical as the  Lagrangian density for it is invariant under a gauge transformation, 
\begin{equation}
\delta_\Lambda h_{ab} = \nabla_a \Lambda_b + \nabla_b \Lambda_a + \pounds_{(\Lambda)} h_{ab}, 
\end{equation}
where

\begin{equation}
\pounds_{(\Lambda)} h_{ab} = \Lambda^c \nabla_c h_{ab} + h_{ac}\nabla_b \Lambda^c + h_{cb} \nabla_a \Lambda^c.
\end{equation}
is the Lie derivative of $h_{ab}$ with respect to the vector field $\Lambda^a$. 

These unphysical degrees of freedom give rise to constraints [17] in the canonical
 quantization  and divergences in the partition function [18] in the path integral quantization.
 So before we can quantize this theory, we need to fix a gauge by adding a gauge fixing term. In order to  ensure unitarity, 
 a ghost term also has to be added. 

Now we first start with the following gauge fixing condition,
\begin{equation}
G[h]_a = (\nabla^b h_{ab} - k  \nabla_a h) = 0,
\end{equation}
where
\begin{equation}
 k \neq 1.
\end{equation}
 For $k=1$, the constraints are not removed and 
the partition function again diverges.  That is why,    $k$ is usually written as $1 + \beta^{-1}$, where $\beta$ is
 an arbitrary finite constant [19].

The  gauge fixing term corresponding to this gauge fixing condition is  given by
\begin{equation}
\mathcal{L}_{gf} = \sqrt{g}\left[ib^a(\nabla^b h_{ab} - k \nabla_a h) + \frac{\alpha}{2}b^a b_a\right],
\end{equation}
and the ghost term is given by  
\begin{equation}
\mathcal{L}_{gh} = i \sqrt{g}\,\overline{c}^a \nabla^b[\nabla_a c_b + \nabla_b c_a - 2 k g_{ab} \nabla_c c^c + \pounds_{(c)} h_{ab} - kg_{ab} g^{cd} \pounds_{(c)} h_{cd}],
\end{equation}
where $\pounds_{(c)} h_{ab}$ is the Lie derivative of $h_{ab}$ with respect to the ghost field $c^a$,
\begin{equation}
\pounds_{(c)} h_{ab}= c^c \nabla_c h_{ab} + h_{ac}\nabla_b c^c + h_{cb} \nabla_a c^c.
\end{equation}
This ghost term can be expressed as 
\begin{equation}
\mathcal{L}_{gh} = \sqrt{g}\,  \overline{c}^a M_{ab} c^b, 
\end{equation}
where $M_{ab}$ is given by
\begin{eqnarray}
M_{ab} &=& i \nabla_c [\delta^c_b \nabla_a + g_{ab} \nabla^c - 2 k \delta^c_a \nabla_b + (\nabla_b h^c_a) - h_{ab} \nabla^c + h_b^c \nabla_a 
 \nonumber \\ && - k g_a^c g^{ef}((\nabla_b h_{ef}) + h_{eb} \nabla_f + h_{fb}\nabla_e)].
\end{eqnarray}
The total Lagrangian density obtained by addition of the original classical Lagrangian, the gauge 
fixing term and the ghost term is invariant under the following BRST transformations, 
\begin{eqnarray}
s \,h_{ab} &=& \nabla_a c_b + \nabla_b c_a + \pounds_{(c)} h_{ab}, \nonumber \\
s \,c^a &=& - c_b \nabla^b c^a, \nonumber \\
s \,\overline{c}^a &=& b^a, \nonumber \\ 
s \,b^a &=&0.
\end{eqnarray}
This total Lagrangian density is also invariant under the following anti-BRST transformations,
\begin{eqnarray}
\overline{s} \,h_{ab} &=& \nabla_a \overline{c}_b + \nabla_b \overline{c}_a + \pounds_{(\overline{c})} h_{ab}, \nonumber \\
\overline{s} \,c^a &=& -b^a - 2 \overline{c}_b \nabla^b c^a, \nonumber \\
\overline{s} \,\overline{c}^a &=& - \overline{c}_b \nabla^b \overline{c}^a,\nonumber \\ 
\overline{s} \,b^a &=& - b^b\nabla_b c^a.
\end{eqnarray}

These BRST and  anti-BRST transformations appear very different.  However, for Yang-Mills theories, the BRST and the anti-BRST transformations
 almost seem to reverse their respective roles, by suitably changing the Nakanishi-Lautrup field [7].

We will now analyse this reversing of the form of   BRST and anti-BRST transformations for perturbative quantum gravity. 
 To do so, we first shift the original Nakanishi-Lautrup field by $2\overline{c}^b\nabla_b c^a$, and then multiply it by $-1$, to get a new Nakanishi-Lautrup field.

Then in terms of this new Nakanishi-Lautrup field the  BRST transformation are given by
\begin{eqnarray}
s\, h_{ab} &=& \nabla_a c_b + \nabla_b c_a + \pounds_{(c)} h_{ab}, \nonumber \\
s \,\overline{c}^a &=& - b^a - 2 \overline{c}_b \nabla^b c^a, \nonumber \\
s \,c^a &=& - c_b \nabla^b c^a,\nonumber \\ 
s \, b^a &=& - b^b\nabla_b \overline{c}^a,
\end{eqnarray}
 and the anti-BRST transformations are given by
\begin{eqnarray}
\overline{s} \,h_{ab} &=& \nabla_a \overline{c}_b + \nabla_b \overline{c}_a + \pounds_{(\overline{c})} h_{ab}, \nonumber \\
\overline{s} \,c^a &=&  b^a, \nonumber \\
\overline{s} \,\overline{c}^a &=& - \overline{c}_b \nabla^b \overline{c}^a, \nonumber \\ 
\overline{s}  \,b^a &=& 0.
\end{eqnarray}

These BRST and anti-BRST transformations  look like the reversed version of the original BRST and anti-BRST transformations.

Both these sets of transformations are nilpotent. In fact they satisfy, 
\begin{equation}
s^2 = \overline{s}^2  = s\overline{s} + \overline{s} s = 0. 
\end{equation}
We can now express the gauge fixing Lagrangian density $\mathcal{L}_g$, which is given by the sum of the gauge fixing term and the ghost term, as follows:
\begin{eqnarray}
\mathcal{L}_g &=&\mathcal{L}_{gf} +\mathcal{L}_{gh} \nonumber \\ &=& i s \sqrt{g} \left[ \overline{c}^a (\nabla^b h_{ab} - k \nabla_a h  
 -  \frac{i \alpha}{2}b_a)\right]\nonumber\\ &=&-i \overline{s}\sqrt{g} \left[ c^a (\nabla^b h_{ab} - k \nabla_a h  
 -  \frac{i\alpha}{2}b_a)\right].
\end{eqnarray}
In a slightly different gauge it can also be written, as follows:
\begin{eqnarray}
\mathcal{L}_g &=& -\frac{i}{2}s \overline{s} \sqrt{g}(h^{ab}h_{ab})+ \frac{i\alpha}{2}\overline{s}\sqrt{g}(b^a c_a)
 \nonumber \\ & =&  \frac{i}{2} \overline{s} s\sqrt{g}(h^{ab}h_{ab})- \frac{i\alpha}{2}s\sqrt{g}(b^a \overline{c}_a).
\end{eqnarray}
Thus the gauge fixing Lagrangian density can be expressed as a total BRST or a total anti-BRST variation. 
\section{Landau Gauge}
In Yang-Mills theories, there is a special gauge called the Landau gauge, in which we can express the gauge fixing Lagrangian density as a 
combination of total BRST and total anti-BRST variations.
 Furthermore, in this gauge the BRST and the anti-BRST variations look similar to each other, with ghosts and anti-ghosts interchanged,
 and the sign of Nakanishi-Lautrup field also changed [13]. 

We will now analyse the BRST and anti-BRST symmetry for perturbative quantum gravity in Landau gauge. 
In  Landau gauge, $\alpha = 0 $,  and so we have 
\begin{eqnarray}
\mathcal{L}_g  &=& i s \sqrt{g}\left[ \overline{c}^a (\nabla^b h_{ab} - k \nabla_a h  ) \right]\nonumber\\ 
&=&-i \overline{s} \sqrt{g}\left[ c^a (\nabla^b h_{ab} - k \nabla_a h  ) \right].
\end{eqnarray}
In a slightly different gauge it can also be written, as follows:
\begin{eqnarray}
\mathcal{L}_g &=&-\frac{i}{2}s \overline{s}\sqrt{g} (h^{ab}h_{ab}) \nonumber \\ & =&  \frac{i}{2} \overline{s} s\sqrt{g}(h^{ab}h_{ab}).
\end{eqnarray}

Thus in Landau gauge the gauge fixing Lagrangian density for perturbative quantum gravity is also expressed as a combination of a total BRST and 
a total anti-BRST variation. 

In Landau gauge the BRST transformations are given by
\begin{eqnarray}
s \,h_{ab} &=& \nabla_a c_b + \nabla_b c_a + \pounds_{(c)} h_{ab}, \nonumber \\
s \,c^a &=& - c_b \nabla^b c^a, \nonumber \\
s \,\overline{c}^a &=& b^a, \nonumber \\ 
s \,b^a &=&0,
\end{eqnarray}
and the anti-BRST transformations are given by  
\begin{eqnarray}
\overline{s} \,h_{ab} &=& \nabla_a \overline{c}_b + \nabla_b \overline{c}_a + \pounds_{(\overline{c})} h_{ab}, \nonumber \\
\overline{s} \,c^a &=& -b^a, \nonumber \\
\overline{s} \,\overline{c}^a &=& - \overline{c}_b \nabla^b \overline{c}^a,\nonumber \\ 
\overline{s} \,b^a &=& 0.
\end{eqnarray}
These transformations look similar to each other, with ghosts and anti-ghosts interchanged, and  the sign of Nakanishi-Lautrup field also changed. 
\section{ Non-Linear Gauges}
Curci-Ferrari Lagrangian density is non-linear in ghosts and anti-ghosts  and thus cannot be obtained directly from the Faddeev-Popov procedure [18]. 
For Yang-Mills theories  in Curci-Ferrari gauge, we can write the gauge fixing Lagrangian density as a combination of total BRST and total
 anti-BRST variations, for any value of $\alpha$ [14]. 
In this section we will express the Lagrangian density for perturbative quantum gravity in Curci-Ferrari gauge as a combination of
a total BRST and a total anti-BRST variation, for any value of $\alpha$. 

The BRST transformations  for perturbative quantum gravity  in Curci-Ferrari gauge are given by
\begin{eqnarray}
s \,h_{ab} &=& \nabla_a c_b + \nabla_b c_a + \pounds_{(c)} h_{ab}, \nonumber \\
s \,c^a &=& - c_b \nabla^b c^a, \nonumber \\
s \,\overline{c}^a &=& b^a - \overline{c}^b\nabla_b c^a, \nonumber \\ 
s \,b^a &=& - b^b\nabla_b c^a -  \overline{c}^b\nabla_b c^d\nabla_d c^a,
\end{eqnarray}
and the anti-BRST transformation for perturbative quantum gravity are given by
\begin{eqnarray}
\overline{s}\, h_{ab} &=& \nabla_a \overline{c}_b + \nabla_b \overline{c}_a + \pounds_{(\overline{c})} h_{ab}, \nonumber \\
\overline{s} \,\overline{c}^a &=& - \overline{c}_b \nabla^b \overline{c}^a, \nonumber \\
\overline{s} \,c^a &=& - b^a - \overline{c}^b\nabla_b c^a, \nonumber \\ 
\overline{s} \,b^a &=& - b^b\nabla_b \overline{c}^a +  c^b\nabla_b\overline{c}^d\nabla_d  \overline{c}^a.
\end{eqnarray}
We can now write a gauge fixing Lagrangian density as a combination of a total BRST and a total  anti-BRST variation, as
\begin{eqnarray}
\mathcal{L'}_g&=& \frac{i}{2}s\overline{s}\sqrt{g}\left[h^{ab}h_{ab} - i \alpha \overline{c}^a c_a \right] \nonumber \\ &=&\frac{-i}{2}\overline{s} s\sqrt{g} \left[h^{ab}h_{ab} - i \alpha \overline{c}^a c_a \right].
\end{eqnarray}

Thus in  the Curci-Ferrari gauge, apart from getting the original Faddeev-Popov part of the gauge fixing Lagrangian,
 we get additional non-linear  contributions proportional to $\sqrt{g}(\overline{c}^b \nabla_b c^a)(\overline{c}^d \nabla_d c_a)$.
 Such terms occur in almost all supersymmetric Yang-Mills theories and  string theory. Furthermore,  such non-linear terms lead to the 
formation of off-diagonal ghost-condensates [20]. However, these ghost-condensates do not give rise to any mass term for the gauge fields. 
This is because the addition of a bare mass term is prevented by the nilpotency of the  BRST and the anti-BRST transformations. However, if we are 
ready to violate the nilpotency of the BRST and the anti-BRST transformations, then we can add such a bare mass term. This has been done for Yang-Mills theories,
 to get a massive  Curci-Ferrari Lagrangian density [15]. Here we will do it for perturbative quantum gravity. 

We can write the massive Curci-Ferrari type of Lagrangian density for perturbative quantum gravity as follows:
 
\begin{eqnarray}
\mathcal{L}_{g}^{ m^2}&=& \frac{i}{2}[s\overline{s}-im^2]\sqrt{g}\left[h^{ab}h_{ab} - i \alpha \overline{c}^a c_a \right] 
\nonumber \\ &=&\frac{i}{2}[-\overline{s} s-im^2] \sqrt{g}\left[h^{ab}h_{ab} - i \alpha \overline{c}^a c_a \right].
\end{eqnarray}
Thus the massive Curci-Ferrari type of Lagrangian density for perturbative quantum gravity contains  contributions proportional to terms like   $\sqrt{g}m^2 g^{ab}g_{ab}$ and  $\sqrt{g}i m^2\alpha\overline{c}^a c_a$. 

This Lagrangian density is invariant under the following BRST transformation,
\begin{eqnarray}
s \,h_{ab} &=& \nabla_a c_b + \nabla_b c_a + \pounds_{(c)} h_{ab}, \nonumber \\
s \,c^a &=& - c_b \nabla^b c^a, \nonumber \\
s \,\overline{c}^a &=& b^a - \overline{c}^b\nabla_b c^a, \nonumber \\ 
s \,b^a &=& i m^2 c^a- b^b\nabla_b c^a -   \overline{c}^b\nabla_b c^d\nabla_d c^a,
\end{eqnarray}
and the following anti-BRST transformations,
\begin{eqnarray}
\overline{s}\, h_{ab} &=& \nabla_a \overline{c}_b + \nabla_b \overline{c}_a + \pounds_{(\overline{c})} h_{ab}, \nonumber \\
\overline{s} \,\overline{c}^a &=& - \overline{c}_b \nabla^b \overline{c}^a, \nonumber \\
\overline{s} \,c^a &=& - b^a - \overline{c}^b\nabla_b c^a, \nonumber \\ 
\overline{s} \,b^a &=& i m^2 \overline{c}^a- b^b\nabla_b  \overline{c}^a +  c^b\nabla_b\overline{c}^d\nabla_d  \overline{c}^a.
\end{eqnarray}
The addition of bare mass term breaks the nilpotency of the BRST and the anti-BRST transformations. The BRST and the anti-BRST transformations now satisfy
\begin{equation}
  s^2 = \overline{s}^2  \sim i m^2.
\end{equation}
However,  in the zero mass limit, the nilpotency of the BRST and the anti-BRST transformations is restored. This breakdown of nilpotency of the BRST and the 
anti-BRST transformations  also leads to  breakdown of the unitarity of the theory. Thus unitarity of this theory is only maintained in the zero mass limit. 
It is possible that in quantum gravity there could be a breakdown of the unitarity [16] and thus it would be interesting to analyse a
 formalism that deals with it.
\section{Conclusion}
In this paper we have  generalized certain results from the Yang-Mills theories in flat spacetime 
to perturbative quantum gravity in curved spacetime. We have shown that  the behaviour of  BRST 
and anti-BRST symmetries for perturbative quantum gravity in arbitrary dimensions, in curved spacetime is similar 
to their behaviour for Yang-Mills theories in four dimensional flat spacetime. Similar to the Yang-Mills theories in flat
spacetime, the BRST and the anti-BRST transformations for perturbative quantum gravity  almost change their respective forms 
by a redefinition of the Nakanishi-Lautrup field, in simple linear gauge with an arbitrary value of $\alpha$.  We   expressed 
the gauge fixing Lagrangian density for perturbative quantum gravity as a combination of  total BRST and total anti-BRST variations, in Landau gauge.
We also expressed the gauge fixing Lagrangian density as a combination of total BRST and total anti-BRST variations, for an arbitrary value of $\alpha$, 
by the adding suitable non-linear terms to it.
Furthermore, the addition of a bare mass term  
violated the nilpotency of the BRST and the anti-BRST transformations, which in turn  violates the  unitarity of the theory. This violation of unitarity could be  
physically relevant in quantum gravity as it is suspected that certain quantum gravitational processes might lead to a breakdown of the unitarity. 
 We stress the fact that all these results were  already known to hold for Yang-Mills theories in flat spacetime and all we have shown here  is that they 
also hold for perturbative quantum gravity in curved spacetime. 
\section{References}
\enumerate 
\item 
S. Weinberg, Gravitation and Cosmology - John Wiley and Sons, New York - (1972)
\item 
C. Becchi, A. Rouet and R. Stora, Annals. Phys. $\bf{98}$, 287 (1976)
\item 
I. Ojima, Prog. Theor. Phys. $\bf{64}$, 625 (1980)
\item 
N. Nakanishi, Prog. Theor. Phys. $\bf{59}$,  972 (1978)
\item 
T. Kugo and I. Ojima, Nucl. Phys. $\bf{B144}$, 234 (1978) 
\item 
K. Nishijima and M. Okawa, Prog. Theor. Phys. $\bf{60}$, 272  (1978) 
\item
N. Nakanishi and I. Ojima, Covariant operator formalism of gauge theories and quantum gravity - World Sci. Lect. Notes. Phys - (1990)
\item
Yoshihisa Kitazawa, Rie Kuriki and Katsumi Shigura, Mod. Phys. Lett. $\bf{A12}$,  1871 (1997)
\item 
E. Benedict, R. Jackiw and  H. J. Lee, Phys. Rev. $\bf{D54 }$, 6213 (1996)
\item 
Friedemann Brandt, Walter Troost and  Antoine Van Proeyen, Nucl. Phys. $\bf{B464}$, 353 (1996) 
\item 
 M. Tahiri, Int. Jou. Theo. Phys. $\bf{35}$, 1572 (1996)
\item 
M. Menaa and M. Tahiri, Phys. Rev. $\bf{D 57}$, 7312 (1998)   
\item 
M. Ghiotti, A. C. Kalloniatis and A.G. Williams. Phys. Lett. $\bf{B628}$,  176 (2005)
\item
L. von Smekal, M. Ghiotti and  A. G. Williams,  Phys. Rev. $\bf{D78}$, 085016 (2008)
\item
G. Curci and R. Ferrari, Phys. Lett. $\bf{B63}$, 91 (1976) 
\item 
S. W. Hawking and J. D. Hayward, Phys. Rev. $\bf{D49}$, 5252 (1994) 
\item 
M. Henneaux  and C. Teitelboim,  Quantization of Gauge Systems. Princeton University Press.  (1992)
\item 
J. T. Mieg, J. Math. Phys. $\bf{21}$, 2834 (1980) 
\item 
A. Higuchi and S. K. Spyros,  Class. Quant. Grav. $\bf{18}$, 4317 (2001) 
\item 
K. I. Kondo and T. Shinohara, Phys. Lett. $\bf{B491}$, 263 (2000) 

\end{document}